\documentclass[preprint,amsmath,amssymb,pre,aps]{revtex4-1}
\usepackage{graphicx}
\usepackage{bm}
\usepackage{amsmath}
\begin{document}
\title{Fluctuation-driven plasticity allows for flexible rewiring of neuronal assemblies}
\author{Federico Devalle, Alex Roxin}
\affiliation{Computational Neuroscience Group, Centre de Recerca Matem\`atica \\Campus de Bellaterra, Edifici C 08193 Bellaterra, Spain.}

\begin{abstract}
Synaptic connections in neuronal circuits are modulated by pre- and
post-synaptic spiking activity.  Heuristic models of this process of
synaptic plasticity can  provide excellent fits to results from
\textit{in-vitro} experiments in which pre- and  post-synaptic spiking
is varied in a controlled fashion. However, the plasticity rules
inferred from  fitting such data are inevitably unstable, in that
given constant pre- and post-synaptic activity the synapse will either
fully  potentiate or depress. This instability can be held in check by
adding additional mechanisms, such  as homeostasis. Here we consider
an alternative scenario in which the plasticity rule itself is stable.
When this is the case, net potentiation or depression only occur when
pre- and post-synaptic activity vary in time, e.g. when driven by
time-varying  inputs. We study how the features of  such inputs shape
the recurrent synaptic connections in models  of neuronal circuits. In
 the case of oscillatory inputs, the resulting structure is  strongly
affected by the phase relationship between drive to different
neurons. In  large networks, distributed phases tend to lead to
hierarchical clustering. Our results may be of relevance for
understanding the effect of sensory-driven inputs, which are by nature
time-varying, on  synaptic plasticity, and hence on learning and memory.  
\end{abstract}
\maketitle
\section{Introduction}
\label{intro}
Learning occurs through changes in the synaptic weights between cells
in neuronal circuits.  Understanding learning therefore requires
working out the rules by which these  changes happen at a single
synapse, and then studying the consequence of this process at the
network level. One major finding regarding the rules underlying
synaptic  plasticity was the observation that the synaptic weight
change could depend on the relative timing of  pre- and post-synaptic
spikes \citep{markram97b,bi98}. Theoretical work has studied how  such spike-timing dependent
plasticity (STDP) can shape recurrent connections in neuronal networks 
\citep{burkitt07,gilson09} and serve to encode fixed  point attractors in
large-scale spiking models \citep{zenke15}. However, like all  Hebbian plasticity
rules, STDP is intrinsically unstable, and leads either to  full saturation of all
synaptic connections, or complete depression, depending on the sign of
the integral of the plasticity window. Essentially, plasticity always
occurs if the  product of pre- and post-synaptic rates is non-zero,
even if constant. Therefore, additional  stabilizing mechanisms are
required in order to allow for the emergence of  non-trivial
connectivity patterns \citep{gutig03,sadeh15,zenke15,zenke17}. A multiplicative 
STDP rule, for which potentiation is progressively weaker the stronger the 
synapse, stabilizes weights but does not readily allow for the emergence of 
non-trivial network structure \citep{vanrossum00,rubin01,morrison07}. 

Here we consider the scenario in which plasticity only occurs in the
presence of time-varying rates. This mode of plasticity seems
particularly relevant for learning given that most salient events
unfold over time and would be expected to result in time-varying
firing rates in the relevant brain circuits. For simplicity we explore
this plasticity regime by  assuming that the integral of the STDP
window is exactly zero, i.e. that there is  a balance between
potentiation and depression when firing rates are constant. This
obviates the need for added stabilizing  mechanisms and allows for an
in-depth analysis. 

We consider both oscillatory as well as noisy drive and develop a
theoretical framework  for pairs of linear firing rate
neurons. Specifically, we make use of the separation of time-scales
between neuronal  and synaptic dynamics to derive self-consistent
evolution equations for the synaptic weights. Analysis of these
equations  reveals a rich phase diagram from which the resulting
connectivity motif can  be predicted depending on the phase difference
in the case of oscillatory drive, or  the correlation and delay in the
case of noisy drive. We also find many regions of  multistability,
meaning that the final connectivity motif will also depend on the
initial configuration of weights. For the case of oscillatory drive we
study the effect of the balanced STDP rule in networks numerically,
and  show that the resulting connectivity matrix can be well predicted
from the  pairwise theory in several relevant cases. 

\section{General two-neuron model}
\label{sec:1}

\subsection{Firing rate formulation for two neurons with forcing}
\label{subsec:1}
We begin by considering the simplest possible scenario of two coupled neurons. We 
model the neuronal activity with a linear firing rate equation, which will allow for 
a complete analysis. The equations are
\begin{equation}
\begin{aligned}
\tau\dot{r}_{1} &= -r_{1}+w_{12}r_{2}+I_{1}(t),\\
\tau\dot{r}_{2} &= -r_{2}+w_{21}r_{1}+I_{2}(t),\label{eq:rate}
\end{aligned}
\end{equation}
where $w_{12}$ and $w_{21}$ are the recurrent synaptic weights. The external 
inputs vary in time; we will consider both oscillatory inputs as well as 
correlated noise sources in subsequent sections. 
\vspace{0.1in} 

\noindent
\fbox{\begin{minipage}{40em}\textbf{Mathematical assumption: Rate dynamics} Neurons are excitable units which can respond in an 
all-or-none manner to inputs. Specifically, if the neuronal membrane potential exceeds a threshold, an 
action potential or ``spike'' is generated, which propagates down the axon and causes the release of 
neurotransmitter onto postsynaptic targets. These excitable dynamics can be modelled 
as a system of coupled nonlinear ordinary differential equations (ODEs)\citep{hodgkin52}, 
or in a simplified manner via one-dimensional ODEs 
with discontinuous threshold and reset conditions\citep{lapicque07,fourcaud03}. 
However, sometimes it is not the details of the subthreshold 
activity of the cell which is of primary interest, and but rather the mean spike rate. The dynamics of this rate can be described in a 
so-called firing rate equation\citep{wilson72}, which is generally heuristic in nature. Here we will explore how variations in the 
spike rate affect plasticity, and hence take this firing rate approach. As a further simplification we take the 
rate dynamics to be linear in Eqs.\ref{eq:rate}.
\end{minipage}}
\vspace{0.1in}

Here we  interpret the  rates in
Eqs.\ref{eq:rate} as the underlying probability for a Poisson
spiking process, and then  apply a so-called spike-timing dependent
plasticity (STDP) rule \citep{kempter99}. That is, in a small time interval 
$\Delta t$, the probability that neuron 1 generates a spike is just 
$r_{1}(t)\Delta t$. Once the spikes have been generated, a presynaptic spike
of neuron $i$ followed by a postsynaptic  spike of neuron $j$ at a
latency $T$ leads to a potentiation of synapse $w_{ji}$ by an amount
$A_{+}e^{-T/\tau_{+}}$ and a depression of synapse $w_{ij}$ by an
amount -$A_{-}e^{-T/\tau_{-}}$. Synapses  are bounded below by zero
and above by a maximum value $w_{max}$.  Whenever there is a new spike
the synaptic weights are updated in this way for all past spike pairs, see 
\citep{pfister06} for an efficient numerical implementation.
This rule, together with Eqs.\ref{eq:rate}, provide a complete model
which can be simulated numerically. Note that 
while we are formally making use of an STDP rule here, the exact 
spike timing plays no role. That is, plasticity is due only to 
dynamics in the rate. Such rates effects appear to dominate over 
contributions due to spike timing in models of STDP when realistic 
input patterns are considered. Specifically, experimental 
protocols have traditionally used highly regular, repeated pairings of 
pre- and post-synaptic activity, and the observed synaptic plasticity has therefore 
been natually attributed to the exact spike timing \citep{markram97b,bi98}. However, theoretical work has 
shown that when the inferred rules are used in the presence of more 
\textit{in-vivo} like spike trains with a high level of irregularity, 
the resulting synaptic plasticty can be accounted for to a large extent just by 
variations in the underlying firing rate \citep{graupner16}.
\vspace{0.1in} 
 
\noindent
\fbox{\begin{minipage}{40em}\textbf{Mathematical assumption: Pairwise STDP} Long-term changes in synaptic strength 
are due to a complex chain of biochemical processes which are set in motion by the in-flux of calcium at the synapse\citep{citri07}. 
Indeed, heuristic models of synaptic plasticity based on the local calcium signal at the synapse can reproduce the diverse 
phenomenology of potentiation and depression oberserved in in-vitro experiments\citep{graupner12}. The calcium signal itself is determined 
by both pre- and post-synaptic spiking, which suggests a phenomenological model based only on the timing of the pre- and 
postsynaptic spikes may provide a good approximate description of the plasticity dynamics\citep{markram97,bi98}. The advantage of such a 
description is that it requires no knowledge of the subthreshold state of the neurons. Such spike-timing dependent plasticity models 
have proven very successful at fitting in-vitro data\citep{pfister06}. Here we use an STDP rule in which changes at the synapse are determined 
solely by pre-post spike pairs. Our self-consistent description of neuronal activity and synaptic plasticity is therefore 
entirely at the level of spiking activity, and hence ignores the details of the subthreshold state of the neurons and calcium 
levels at the synapse. 
\end{minipage}}
\vspace{0.1in}

If the amplitude of potentiations and depressions is small compared to
the  maximum synaptic strength $w_{max}$, then the evolution of the
synaptic weights can  be approximated by the following integrals
\citep{kempter99} 
\begin{eqnarray}
\dot{w}_{ij} &=&
-A_{-}\int_{-\infty}^{0}dTe^{T/\tau_{-}}r_{j}(t)r_{i}(t+T)\nonumber\\ &&+A_{+}\int_{0}^{\infty}dTe^{-T/\tau_{+}}r_{i}(t)r_{j}(t-T).\label{eq:wij_ns}
\end{eqnarray}
The first integral includes the contribution of all spike pairs
leading to  depression of the synapse $w_{ij}$. Specifically, given
Poisson processes, the  probability of the spike pair in which cell
$i$ has spiked in a small interval  around time $t$ and cell $j$ has
spiked previously at a latency $T$ is just  $r_{i}(t)r_{j}(t-T)$
($T>0$). The integral sums up spike pairs at all possible latencies up
until  the current time. The second integral is analogous, but for
spike pairs which lead to  potentiation of the synapse. When the 
neuronal dynamics is stationary, the integrals can be written in the 
compact form
\begin{equation}
\dot{w}_{ij} = \int_{-\infty}^{\infty}dTA(T)r_{j}(t)r_{i}(t+T), \label{eq:wij}
\end{equation} 
where $A(T) = A_{+}e^{-T/\tau_{+}}$ for $T>0$ and $-A_{-}e^{T/\tau_{-}}$ for $T<0$, see 
\textit{Appendix A} for a detailed explanation. Eqs.\ref{eq:wij} together with the rate equations Eqs.\ref{eq:rate} 
consitute a self-consistent approximation to the full model, and which is 
ammenable to analysis. 

\subsection{Asymptotic approximation for slow plasticity}
Eqs.\ref{eq:rate} and \ref{eq:wij} cannot be solved for directly. The
reason is the  presence of quadratic nonlinearities of the
firing rate variables in the integrals in  Eqs.\ref{eq:wij}. However,
we can take advantage of the slowness of synaptic plasticity compared
to the firng rate dynamics to derive an approximate system of
equations which can be solved  exactly. Specifically, we assume that
the amplitude of potentiation and that of depression are  small and
formalize this by replacing the kernel $A(T) = \epsilon \bar{A}(T)$ in
Eqs.\ref{eq:wij}.  We also define a new, slow time $t_{s} = \epsilon
t$ and allow the rates and the  synaptic weights to evolve
both on a fast as well as on a slow timescale. We expand the rates and weights
in orders of $\epsilon$ and find that the leading order solution,
where $(r_{1},r_{2},w_{12},w_{21}) =
(r_{1}^{0},r_{2}^{0},w_{12}^{0},w_{21}^{0})+\mathcal{O}(\epsilon )$,
obeys the following coupled equations
\begin{equation}
\begin{aligned}
\tau\partial_{t}r_{1}^{0} &=
-r_{1}^{0}+w_{12}^{0}r_{2}^{0}+I_{1}(t),\\ \tau\partial_{t}r_{2}^{0}
&=
-r_{2}^{0}+w_{21}^{0}r_{1}^{0}+I_{2}(t),\\ \partial_{t_{s}}w_{12}^{0}
&= \int
dt\int_{-\infty}^{\infty}dT\bar{A}(T)r_{2}^{0}(t)r_{1}^{0}(t+T),\\ \partial_{t_{s}}w_{21}^{0}
&= \int
dt\int_{-\infty}^{\infty}dT\bar{A}(T)r_{1}^{0}(t)r_{2}^{0}(t+T),\\\label{eqs:asymp_leading}
\end{aligned}
\end{equation}
see \textit{Appendix A} for details of the derivation. In Eqs.\ref{eqs:asymp_leading}
there is a formal separation of the timescale  of evolution of the
rates from that of the synaptic weights, which is much slower. In
fact, the synaptic weights  are only a function of the slow-time and
hence can be treated as constants in the first two equations, allowing
one to solve for the rates using techniques from linear algebra. The
integrals can then be formally evaluated,  yielding self-consistent
evolution equations for the synaptic weights alone. The integrals over
the fast time  are performed over an appropriate time window,
e.g. over one period of oscillation for oscillatory drive.
\vspace{0.1in} 
 
\noindent
\fbox{\begin{minipage}{40em}\textbf{Mathematical assumption: Separation of timescales} Neuronal 
membrane time constants generally range from milliseconds to tens of milliseconds. On the other 
hand, the time course of plasticity as gleaned from in-vitro and in-vivo studies can be much slower. 
Specifically, in in-vitro protocols for the induction of plasticity via STDP, repeated pairings 
are required in order to observe a change in the synaptic efficacy\citep{bi98}. On the other hand, rapid plasticity can 
also be induced experimentally through burst protocols\citep{bliss73}, and has been inferred from fast remapping 
of place cell activity in the hippocampus\citep{priestly22}. Certainly, the formation of episodic 
memory requires plasticity to be fast enough for one-shot learning. 
Our assumption of a separation of timescales therefore means we are modelling 
slow changes in cell responses, perhaps related to perceptual learning. 
\end{minipage}}
\vspace{0.1in}

\section{Oscillatory Drive}
\label{section:osc}

We first consider the case of oscillatory drive with frequency
$\omega$ and  phase difference $\phi $. Specifically, we take
$I_{1}(t) = Ie^{i\omega t+i\phi_{1}}$ and  $I_{2}(t) = Ie^{i\omega
  t+i\phi_{2}}$. The physiological firing rates are given by the real
parts of $r_{1}$ and $r_{2}$, which are then also used to
calculate the weights self-consistently in  Eqs.\ref{eqs:asymp_leading}, yielding
\begin{equation}
\begin{aligned}
\dot{w}_{12} &= \frac{|R|^{2}}{4}\Bigg[\Big(w_{12}+w_{21}+(w_{12}w_{21}+1+\tau^{2}\omega^{2})\cos{\phi}\Big)
\Big(\tilde{A}_{+}(\omega)-\tilde{A}_{-}(\omega)\Big)\\
&+\Big(\tau\omega (w_{12}-w_{21})+(1+\tau^{2}\omega^{2}-w_{12}w_{21})\sin{\phi}\Big)
\Big(\tilde{A}_{+}(\omega)\tau_{+}\omega +\tilde{A}_{-}(\omega)\tau_{-}\omega \Big)\Bigg],\\
\dot{w}_{21} &= \frac{|R|^{2}}{4}\Bigg[\Big(w_{12}+w_{21}+(w_{12}w_{21}+1+\tau^{2}\omega^{2})\cos{\phi}\Big)
\Big(\tilde{A}_{+}(\omega)-\tilde{A}_{-}(\omega)\Big)\\
&-\Big(\tau\omega (w_{12}-w_{21})+(1+\tau^{2}\omega^{2}-w_{12}w_{21})\sin{\phi}\Big)
\Big(\tilde{A}_{+}(\omega)\tau_{+}\omega +\tilde{A}_{-}(\omega)\tau_{-}\omega \Big)\Bigg],\label{eq:ws}
\end{aligned}
\end{equation}

where $\tilde{A}_{+} =
\frac{A_{+}\tau_{+}}{1+\tau_{+}^{2}\omega^{2}}$, $\phi =
\phi_{2}-\phi_{1}$  and we have left off the superscript 0 for
simplicity. 
\vspace{0.1in}

\noindent
\fbox{\begin{minipage}{40em}\textbf{Physiological assumption: oscillatory drive} Oscillations are ubiquitous in the 
brain, although it remains unknown for the most part what their functional role might be. It has been 
hypothesized that the phase relationship between neuronal ensembles oscillating at the same frequency may 
influence their communication\citep{fries05}. Here we explore that possibility that this phase relationship may play a role 
in influencing the directionality and degree of synaptic plasticity. 
\end{minipage}}
\vspace{0.1in}

\begin{figure}
\includegraphics[width=0.55\textwidth]{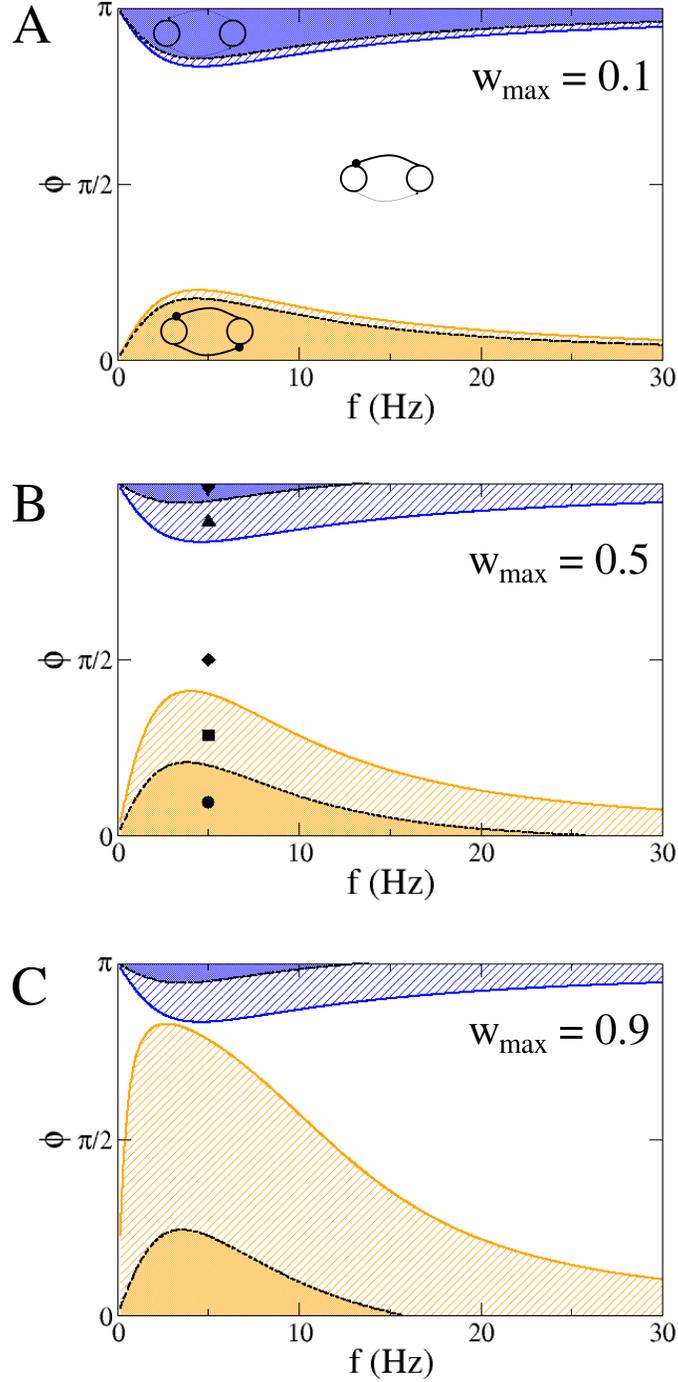}
\caption{Phase planes of the synaptic dynamics in Eqs.\ref{eq:ws} as a
  function  of the phase difference $\phi$ and frequency $f =
  \omega/(2\pi)$ of the forcing, and for different  values of
  $w_{max}$. A. Phase plane for $w_{max} = 0.1$. Bidirectional,
  unidirectional and unconnected  motifs are stable in the orange,
  white and blue regions respectively. The unidirectional motif  is
  stable in the region demarked by the two dashed lines; hence the
  system is bistable in the hatched regions.  B. Phase plane for
  $w_{max} = 0.5$. The symbols indicate the parameter values used for
  the  simulations in Fig.\ref{fig:traces}. C. Phase plane for
  $w_{max} = 0.9$. Other parameters:  $\tau = 10$ms, $\tau_{+} =
  20$ms, $\tau_{-} = 60$ms, $A_{+} = 0.001$, $A_{-} =
  A_{+}\tau_{+}/\tau_{-}$,  $I_{0} = 30$Hz, $I =
  20$Hz.}\label{fig:phaseplane}
\end{figure}

An analysis of Eqs.\ref{eq:ws} reveals that there are no
fixed point solutions for  $w_{12},w_{21}\ge 0$. However, by studying the sign of 
the right hand side in Eqs.\ref{eq:ws} at the boundaries of the allowable domains, we can find 
stable solutions. For example, the fully potentiated  solution
(bidirectional motif) is stable if $\dot{w}_{12},\dot{w}_{21}>0$ for
$(w_{12},w_{21}) = (w_{max},w_{max})$. This condition is clearly
satisfied for $\phi = 0$ as long as
$\tilde{A}_{+}(\omega)-\tilde{A}_{-}(\omega)>0$, which holds when
potentiation dominates at short latencies. On the other hand, when
the $\phi = \pi$, $\dot{w}_{12},\dot{w}_{21}<0$, meaning that there is a critical value of the phase for
which  the fully potentiated solution becomes unstable. In a plane of
the phase versus the  frequency of the forcing, there is therefore a
curve below which the potentiated solution is stable, see the orange
line in Fig.\ref{fig:phaseplane}. An  analogous argument can be made
for the fully depressed solution (unconnected motif), which in this
case is stable above a critical curve, see the blue line in
Fig.\ref{fig:phaseplane}.  Finally, the solution for which one synapse
is fully potentiated and the other fully depressed (unidirectional
motif) is  stable between the dashed, black lines in
Fig.\ref{fig:phaseplane}.  Note that there is a region of bistability
between the unidirectional motif and the bidirectional (unconnected)
motif, indicated by the  orange (blue) hatching in
Fig.\ref{fig:phaseplane}. 

\begin{figure}
\includegraphics[width=0.8\textwidth]{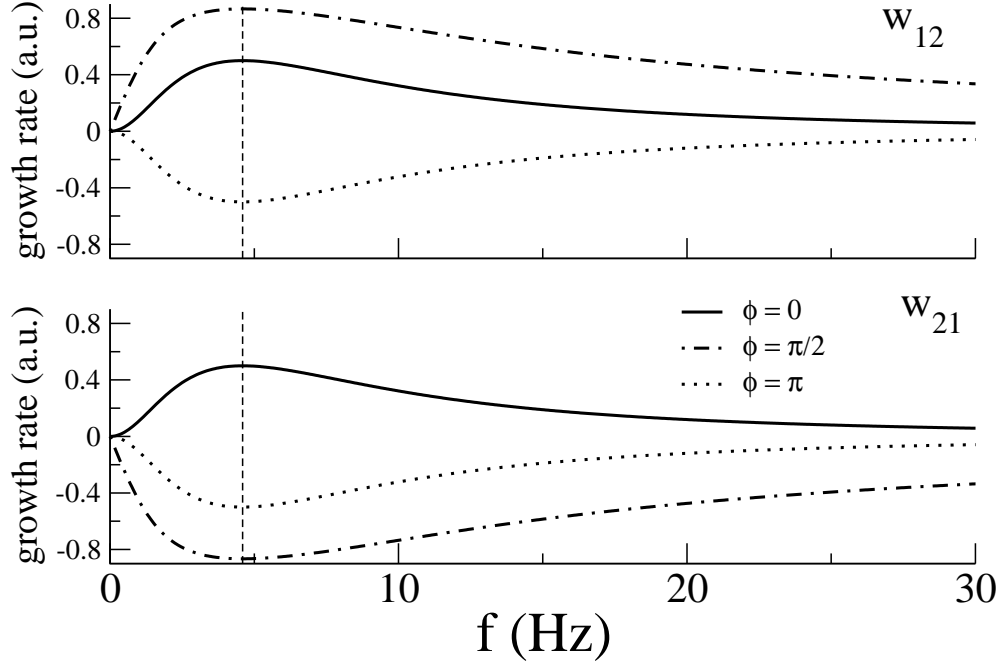}
\caption{Growth rates proportional to 
$\dot{w}_{12}$ and $\dot{w}_{21}$ for $(w_{12},w_{21}) = (0,0)$ as a function of the forcing frequency and for 
three different phases $\phi$. Note that the optimal forcing frequency is independent of phase and equal to 
$f_{opt} = 1/2\pi\sqrt{\tau_{+}\tau_{-}}$ (vertical dashed line). All other parameters are as in Fig.\ref{fig:phaseplane}}\label{fig:growth}
\end{figure}

The phase planes shown in Fig.\ref{fig:phaseplane} show a clear
resonance for values of a critical forcing frequency, approximately
around 5Hz in this case. Specifically, the fully potentiated and fully
depressed  states are stable over a much wider range of forcing phases
in this regime. Additionally, and as  illustrated in
Fig.\ref{fig:growth}, the rate of change of the synaptic weights is
also maximal around  this value of the frequency, and decreases to
zero for zero frequency and in the limit of high  frequencies. Both
effects can be understood as the  interaction of the forcing frequency
with the window of plasticity, i.e. there is a "best" frequency which
maximizes the integral in Eq.\ref{eq:wij}.
Precisely this resonance mechanism  has been invoked to explain the
role of theta oscillations in driving plasticity in  rodent
hippocampus \citep{theodoni18}.  This optimal frequency can be 
found by taking the derivative of the growth rate as a function of the 
forcing frequency. Doing so in the case where $w_{12} = w_{21} = 0$ leads to 
the simple relation $f_{opt} = 1/(2\pi\sqrt{\tau_{+}\tau_{-}})$, independent 
of forcing frquency, see vertical dashed line in Fig.\ref{fig:growth}.

\begin{figure}
\centerline{\includegraphics[width=0.7\textwidth]{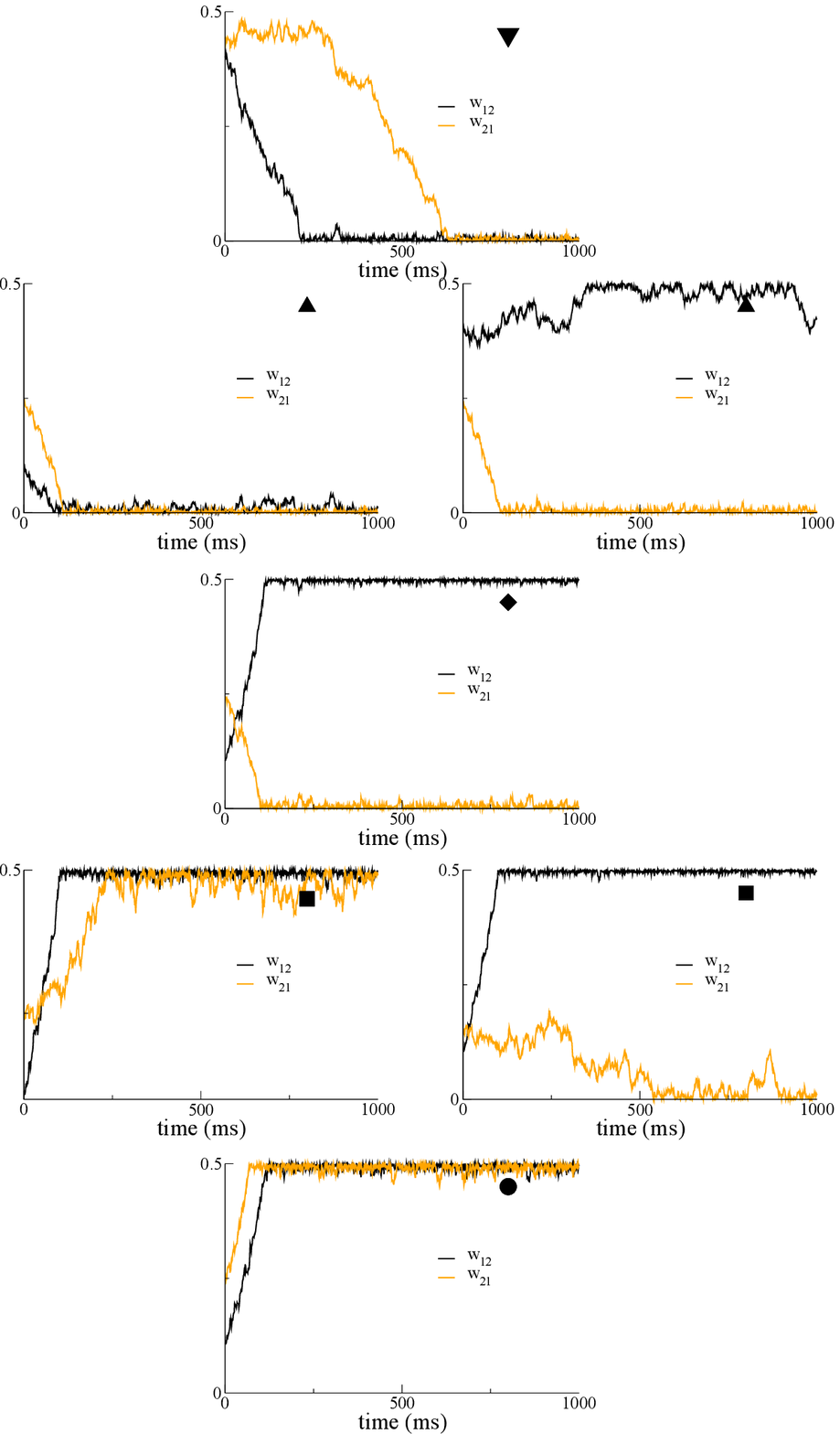}}
\caption{Sample synaptic dynamics for parameter values used in
  Fig.\ref{fig:phaseplane}B. The forcing frequency  is fixed at $f =
  5$Hz and the phase is indicated by the corresponding symbol in
  Fig.\ref{fig:phaseplane}. The different final  states in the traces
  for identical parameter values are due to different initial
  conditions.}\label{fig:traces}
\end{figure}

Numerical simulations of the full model agree well with the analysis
of the reduced system.  Specifically, given a fixed forcing frequency,
there is a range of phase-differences near the in-phase forcing which
lead to both synapses potentiating (PP) (orange region in
Fig.\ref{fig:phaseplane}). Fig.\ref{fig:traces} shows an example of
the synaptic dynamics in this region for  forcing frequency and phase
indicated by the black circle in Fig.\ref{fig:phaseplane}B. For
slightly larger phase differences two sets of  synaptic weights can
coexist depending on initial conditions: PP or one potentiated and the
other depressed (DP) (orange hatched region in
Fig.\ref{fig:phaseplane}), see example simulations in
Fig.\ref{fig:traces} for the parameters given by the black square. In
an intermediate range of phase differences between in-phase and
anti-phase, only uni-directional connectivity emerges (DP) (white
region in Fig.\ref{fig:phaseplane}), see  an example simulation in
Fig.\ref{fig:traces} for the parameters given by the black
diamond. Finally, close to an anti-phase forcing there is a region  of
bistability between (DP) and a fully disconnected motif (DD), followed
by a region in which only the DD solution is stable (blue hatched
region and  solid blue region in Fig.\ref{fig:phaseplane}
respectively), see sample simulations for the black up- and
down-triangles in Fig.\ref{fig:traces}  respectively.

\subsection{Oscillation-driven plasticity in networks}

\begin{figure}
\centerline{\includegraphics[width=0.7\textwidth]{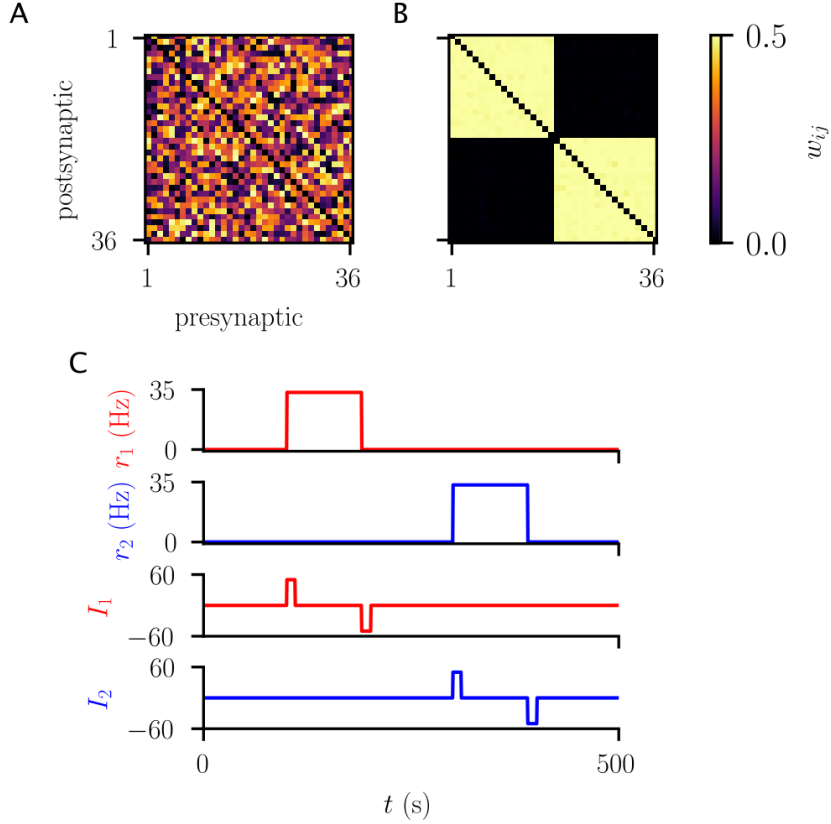}}
\caption{Driving two populations of neurons out-of-phase leads to the
  formation of two disconnected clusters. A. The initial synaptic
  weight distribution. Weights are uniformly  distributed between 0
  and 0.5. B. The synaptic weight matrix at long times after
  oscillatory forcing of two groups of 18 neurons each, with  a phase
  difference of $\pi $ between the two groups. Simulation time is
  $10^{6}$ms. Parameters:  $I_{0} = 5$, $\alpha^{-1} = 0.03$, $I =
  I_{osc}\cos{\omega t}$,  $I_{osc} = 30$. Coupling normalization is
  $K = N_{cl}$.  C. After the learning period the two neuronal
  populations exhibit bistability due to the strong recurrent
  connections.  For these simulations the neurons are taken to have a
  Heaviside transfer function.}\label{fig:two_clusters}
\end{figure}

In the previous section we derived evolution equations for the
synaptic weights for a single pair of neurons. We showed that these
equations can admit several different stable configurations of weights
depending mainly on the phase difference of the forcing, while the
frequency chiefly affected the  learning rates. We now study the
evolution of synaptic weights in a network of an arbitrary number of
neurons. Once again we drive all of the neurons with an oscillatory
forcing at a fixed frequency, and with a phase which can differ from
cell to cell. The resulting synaptic weight matrix is expected to
depend on the precise choice of phases. In principle we can make use
of the same theory to derive a set of coupled ODEs for the synaptic
weights, akin to Eqs.\ref{eq:ws}, see \textit{Appendix B} for details. This leads to N(N-1) coupled
equations for  a network of size N. However, there are some simple
cases for which the resulting synaptic weight matrix can be
straightforwardly predicted from the  pairwise theory. The simplest
example is that of a network in which half of the neurons are driven
at one phase, and the other half at a different phase.  In this case,
neurons within a cluster have zero phase difference between them,
leading to the strengthening of recurrent  connections, while neurons
from different clusters will have their connection shaped according to
the given phase difference.  For a phase difference of $\pi $, the
between-cluster connections decay to zero, leading to the formation of
two  unconnected clusters, as shown in Fig.\ref{fig:two_clusters}B. 

Note that, even though the theory was developed for linear firing rate
neurons, it nonetheless correctly predicts the final state of the
weight matrix even for nonlinear rate neurons, at least in this simple
case. In fact, for the simulations shown in
Fig.\ref{fig:two_clusters}, the neuronal transfer function is taken to
be a Heaviside function, see \textit{Appendix B} for details. Given this choice, the clusters, once formed,
can exhibit bistability, see Fig.\ref{fig:two_clusters}C.

\begin{figure}
\centerline{\includegraphics[width=0.7\textwidth]{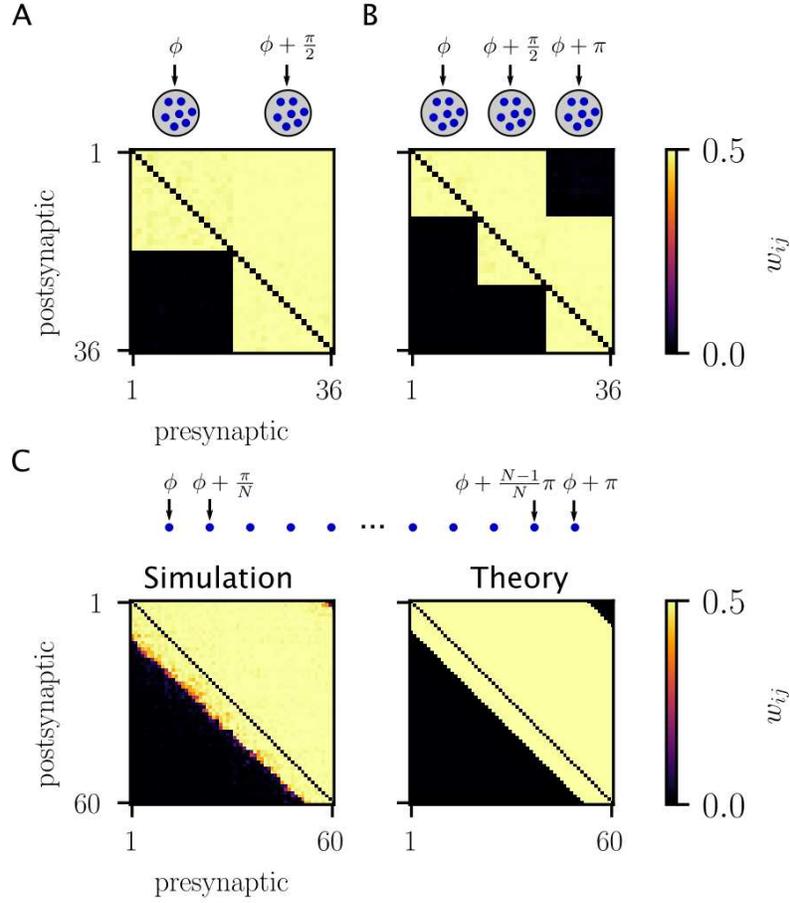}}
\caption{Network connectivity depends on the distribution of phases
  and is well-predicted from pairwise theory. A. A two-cluster
  network with a phase difference of $\pi/2$. B. A three-cluster
  network with phases $\phi_{1} = \phi$, $\phi_{2}=\phi+\pi/2$  and
  $\phi_{3}=\phi+\pi$. C. A network with phases distributed uniformly
  between 0 and $\pi$. Other parameters are the same as in
  Fig.\ref{fig:two_clusters}}\label{fig:networks}
\end{figure}

The applicability of the pairwise theory to networks in which the
forcing is clustered is further illustrated in
Fig.\ref{fig:networks}. Specifically, with  a phase difference of
$\pi/2$ in a two-cluster network, we expect the cross-cluster
connectivity to be unidirectional, from the leader to the follower.
This is indeed what is found in simulation, see
Fig.\ref{fig:networks}A. For the case of three clusters, in which
clusters 2 and 3 have a phase  difference of $\pi/2$ and $\pi$ with
cluster 1 respectively, Fig.\ref{fig:networks}B shows that the same
unidirectional motif is found between clusters 2 and 1, and 3 and 2,
while 1 and 3 become uncoupled, all as predicted from the pairwise
theory. 

The theory can furthermore be extended to the case  in which
the phase difference is distributed uniformly across the network, from
0 to $\pi$. In this case the interaction between any 
pair of neurons depends mainly on the weights between those two neurons 
because the influence of the rest of the network is close to zero, see 
\textit{Appendix B} for details. In this case, neurons with similar phases are expected to
form strong recurrent connections, while sufficiently different phase
difference will lead to leader-follower unidirectional motifs, and
phase differences  near $\pi $ will lead to complete
uncoupling. Numerical simulations show that the resulting synaptic
weight matrix is, in fact, very close to that predicted  from the pure
pairwise theory, see Fig.\ref{fig:networks}C.

\section{Noisy Drive}
\label{sec:noise}

In this section we consider a pair of neurons driven by noisy
inputs. The inputs are
\begin{equation}
\begin{aligned}
I_{1}(t) &=
I_{0}+\sqrt{1-|c|}\xi_{1}(t)+\mathrm{sgn}(c)\sqrt{|c|}\xi_{c}(t),\\ I_{2}(t)
&= I_{0}+\sqrt{1-|c|}\xi_{2}(t)+\sqrt{|c|}\xi_{c}(t-D),
\end{aligned}
\end{equation}
where $\langle\xi_{i}(t)\rangle = 0$ and
$\langle\xi_{i}(t)\xi_{j}(t^{'})\rangle =
\sigma^{2}\delta_{ij}\delta(t-t^{'})$,  and $i,j =
\{1,2,c\}$. Therefore, each neuron receives drive from one independent
noise source, and one common  noise source. The correlation in the
drive between the two noise sources is $c$, and the input to neuron 2
is lagged by an amount $D$.  Given this input, we can calculate the
self-consistent dynamics for the synaptic weights, as before, assuming
that the  impact of each spike pair is weak compared to the dynamic
range of the synapse. The equations are

\small
\begin{equation}
\begin{aligned}
\dot{w}_{12} &= \frac{\sigma^{2}}{8}\Bigg[\Big(\sqrt{\frac{w_{21}}{w_{12}}}+\sqrt{\frac{w_{12}}{w_{21}}}\Big)\Big(\frac{F(\tau_{s},\tau_{s})}{1-\sqrt{w_{12}w_{21}}}
-\frac{F(\tau_{f},\tau_{f})}{1+\sqrt{w_{12}w_{21}}}\Big)-\Big(\sqrt{\frac{w_{21}}{w_{12}}}-\sqrt{\frac{w_{12}}{w_{21}}}\Big)\Big(F(\tau_{s},\tau_{f})-F(\tau_{f},\tau_{s})\Big)\Bigg]\\
&+c\frac{\sigma^{2}}{8}\Bigg[\frac{e^{-D/\tau_{f}}}{1+\sqrt{w_{12}w_{21}}}\Big(G_{-}(\tau_{f},\tau_{f})+G_{+}(\tau_{f},\tau_{f})\Big)+\frac{e^{-D/\tau_{s}}}{1-\sqrt{w_{12}w_{21}}}
\Big(G_{-}(\tau_{s},\tau_{s})+G_{+}(\tau_{s},\tau_{s})\Big)\\
&+e^{-D/\tau_{s}}\Big(G_{-}(\tau_{s},\tau_{f})-G_{+}(\tau_{s},\tau_{f})\Big)+e^{-D/\tau_{f}}\Big(G_{-}(\tau_{f},\tau_{s})-G_{+}(\tau_{f},\tau_{s})\Big)\Bigg].\\
\dot{w}_{21} &= \frac{\sigma^{2}}{8}\Bigg[\Big(\sqrt{\frac{w_{21}}{w_{12}}}+\sqrt{\frac{w_{12}}{w_{21}}}\Big)\Big(\frac{F(\tau_{s},\tau_{s})}{1-\sqrt{w_{12}w_{21}}}
-\frac{F(\tau_{f},\tau_{f})}{1+\sqrt{w_{12}w_{21}}}\Big)+\Big(\sqrt{\frac{w_{21}}{w_{12}}}-\sqrt{\frac{w_{12}}{w_{21}}}\Big)\Big(F(\tau_{s},\tau_{f})-F(\tau_{f},\tau_{s})\Big)\Bigg]\\
&+c\frac{\sigma^{2}}{8}\Bigg[\frac{e^{-D/\tau_{f}}}{1+\sqrt{w_{12}w_{21}}}\Big(G_{-}(\tau_{f},\tau_{f})+G_{+}(\tau_{f},\tau_{f})\Big)+\frac{e^{-D/\tau_{s}}}{1-\sqrt{w_{12}w_{21}}}
\Big(G_{-}(\tau_{s},\tau_{s})+G_{+}(\tau_{s},\tau_{s})\Big)\\
&-e^{-D/\tau_{s}}\Big(G_{-}(\tau_{s},\tau_{f})-G_{+}(\tau_{s},\tau_{f})\Big)-e^{-D/\tau_{f}}\Big(G_{-}(\tau_{f},\tau_{s})-G_{+}(\tau_{f},\tau_{s})\Big)\Bigg],\label{eq:ws_noise}
\end{aligned}
\end{equation}  
\normalsize
where
\begin{eqnarray}
F(\tau_{1},\tau_{2}) &=& \frac{A_{+}\tau_{+}\tau_{1}}{\tau_{1}+\tau_{+}}-\frac{A_{-}\tau_{-}\tau_{2}}{\tau_{2}+\tau_{-}},\\
G_{+}(\tau_{1},\tau_{2}) &=& \frac{A_{+}\tau_{+}\tau_{1}}{\tau_{1}+\tau_{+}}e^{-\frac{D}{\tau_{+}}}
+\frac{A_{+}\tau_{+}\tau_{2}}{\tau_{+}-\tau_{2}}(e^{-\frac{D}{\tau_{+}}}-e^{-\frac{D}{\tau_{2}}})\nonumber\\
&&-\frac{A_{-}\tau_{-}\tau_{2}}{\tau_{2}+\tau_{-}}e^{-\frac{D}{\tau_{2}}},\\
G_{-}(\tau_{1},\tau_{2}) &=& \frac{A_{+}\tau_{+}\tau_{1}}{\tau_{1}+\tau_{+}}e^{-\frac{D}{\tau_{1}}}
+\frac{A_{+}\tau_{-}\tau_{1}}{\tau_{-}-\tau_{1}}(e^{-\frac{D}{\tau_{1}}}-e^{-\frac{D}{\tau_{-}}})\nonumber\\
&&-\frac{A_{-}\tau_{-}\tau_{2}}{\tau_{2}+\tau_{-}}e^{-\frac{D}{\tau_{-}}}.
\end{eqnarray}
Note that $\lim_{D\to 0}{G_{+}} = \lim_{D\to 0}G_{-} = F$. It may appear from Eqs.\ref{eq:ws_noise} that the dynamics 
is singular for $w_{12}\to 0 $ or $w_{21}\to 0$, but these limits are, in fact, well defined, see 
\textit{Appendix C} for details. 
\vspace{0.1in}

\noindent
\fbox{\begin{minipage}{40em}\textbf{Physiological assumption: Correlated, noisy drive} Neuronal activity reflects in part sensory input 
and in part the internal state of the animal. Many neurons respond selectively to particular features of a sensory stimulus, for example 
the orientation of bar\citep{hubel62}, or the direction of motion of an object\citep{mikami86}. In the face of a generic time-varying sensory input, neurons with 
similar feature selectvities in a given cortical area will likely receive correlated inputs, e.g. they may share presynaptic inputs. 
Similarly, neurons with quite different feature selectivities may receive nearly uncorrelated or even negatively correlated inputs. 
In this section we investigate how the recurrent connectivity between such neurons is affected by the degree of correlation in their 
inputs. We choose Gaussian white-noise processes with a given correlation for simplicity.
\end{minipage}}
\vspace{0.1in}

\begin{figure}
\includegraphics[width=0.6\textwidth]{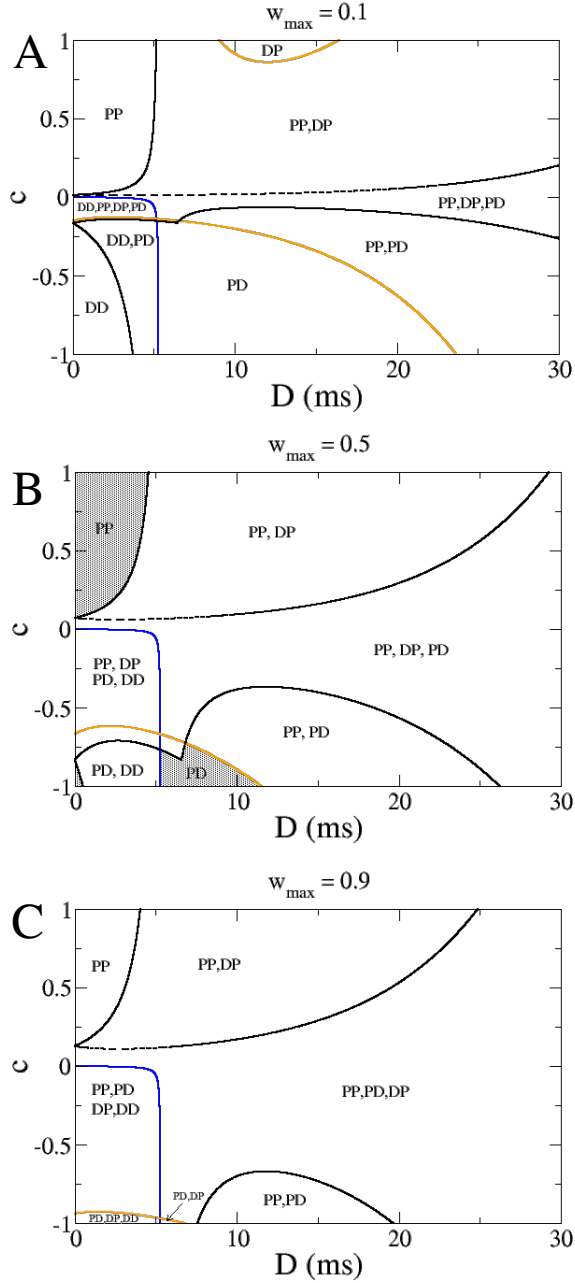}
\caption{Phase planes of synaptic states for two noise-driven neurons
  as a function of the correlation and delay in the shared noisy
  drive.  A. Phase plane for the case $w_{max} = 0.1$. B. Phase plane
  for the case $w_{max} = 0.5$. C. Phase plane for the case $w_{max} =
  0.9$.  Other parameters: $\tau = 10$ms, $\sigma = 5$, $I_{0} = 10$,
  $A_{+} = 0.001$, $\tau_{+} = 20$ms, $\tau_{-} = 60$ms, $A_{-} =
  A_{+}\tau_{+}/\tau_{-}$.  The shaded regions in B. indicate a single
  stable state.}\label{fig:phaseplanes_noise}
\end{figure}

As before, there are no fixed point solutions of
Eqs.\ref{eq:ws_noise}.  However, the equations together with the condition
that $0\le w_{ij}\le w_{max}$, can be used to determine which synaptic
states are stable, as  in the previous section.
Fig.\ref{fig:phaseplanes_noise} shows sample phase planes as a
function of the correlation and delay in the common noisy drive, and
for  three values of $w_{max}$. In all cases, the fully potentiated
state (PP) and the asymmetric state (DP) are favored for positive
correlations. The state  (DP) corresponds to a potentiated connection
from 1 to 2, which occurs when the delay is sufficiently long (2
follows 1 here). For negative correlations  the phase diagram is more
complex, allowing for up to eight distinct regions. In general, the
fully depressed state is stable when correlations are negative  and
the delay is not too large. On the other hand, the asymmetric state
(PD) is stable everywhere for negative $c$.    \normalsize

\begin{figure}
\includegraphics[width=0.7\textwidth]{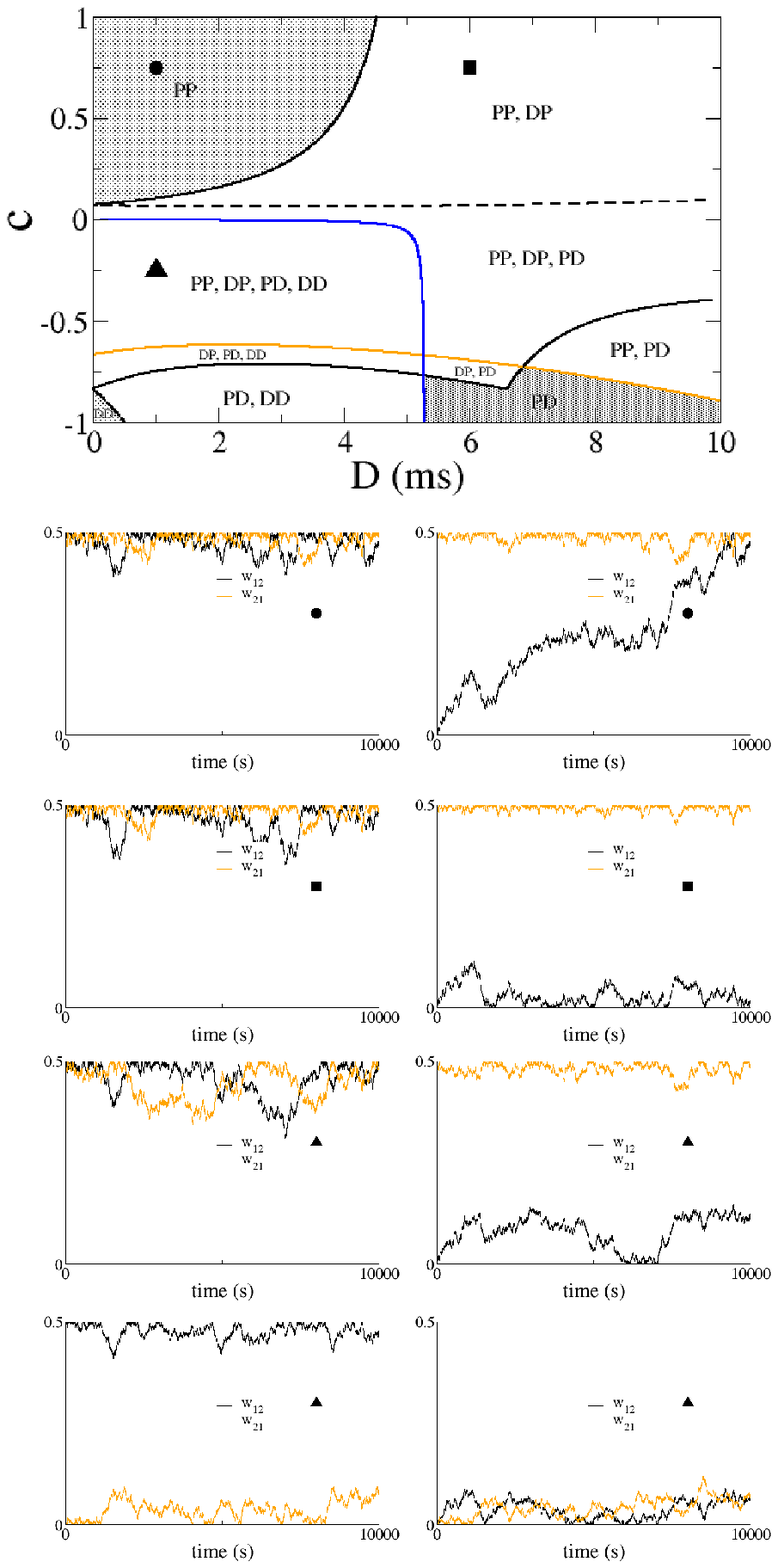}
\caption{A blow-up of the phase plane from
  Fig.\ref{fig:phaseplanes_noise}B, with illustrative numerical
  simluations. The symbols indicate  parameter values used for the
  numerical simulations.}\label{fig:phaseplane_w_traces}
\end{figure}

Fig.\ref{fig:phaseplane_w_traces} shows a detail of the phase diagram
in Fig.\ref{fig:phaseplanes_noise}B with symbols indicating  parameter
values use to confirm the analytical results through several
illustrative numerical simulations. 


\section{Discussion}
We have derived a set of equations for the joint evolution of neuronal
activity and of the synaptic weights between pairs of neurons assuming
a separation of time-scales between the two processes,
Eqs.\ref{eqs:asymp_leading}. The resulting equations can be solved
analytically in the case of linear firing rate neurons, and reduce to
a set of coupled ODEs for the synaptic weights alone. For periodic 
and  noisy forcing the resulting equations are Eqs.\ref{eq:ws} and
Eqs.\ref{eq:ws_noise} respectively. The plasticity rule we have chosen
is  formally a spike-timing dependent rule (STDP), although the fact
that we generate spikes as a Poisson process insures that the actual
spike timing plays no role here. Rather it is only variations in the
underlying rates for the Poisson processes which can lead to
plasticity in our model. This appears to  be the dominant factor in
shaping plasticity given realistic spike trains, even when spiking
correlations are taken into account \citep{graupner16}.  We
additionally assume that the plasticity rule is balanced, namely that
the integral over the STDP window is identically zero. If this is not
the  case, then the synaptic weights between neurons with non-zero
rates would always grow or decay, depending on the sign of the
integral.  This implies that in network simulations the final synaptic
weight matrix will always saturate or decay to zero, ruling out the
emergence of  any non-trivial structure. Additional mechanisms, such
as homeostasis, are needed in this case in order to avoid saturation
\citep{zenke15}.  On the other hand, with the balance assumption only
time-variations in the firing rates can drive
plasticity. Specifically, the change in a given synaptic weight
depends on the covariance of the pre- and post-synaptic firing
weights, multiplied by the STDP window \citep{theodoni18}. 

We also note that our equations do not take into account the variability 
arising due to the stochasticity of firing, but rather only the mean 
Poisson rates. In numerical simulations of the full, 
stochastic system, one observes fluctuations which may momentarily drive the 
synaptic weight away from its stable meanfield value (this is clearly 
seen in Fig.\ref{fig:phaseplane_w_traces}) or even cause transitions 
between stable states. These effects are not captured by Eqs.\ref{eq:ws} 
or Eqs.\ref{eq:ws_noise}.

\textit{Oscillatory drive:} In the case of two periodically forced
neurons, the  resulting connectivity motif depends solely on the phase
difference, while the  time it takes for the connectivity to reach its
steady state depends strongly on  the forcing frequency. 
When potentiation dominates at short latencies then small phase
differences lead to  a fully potentiated motif. Larger phase
differences generate a unidirectional motif in which  the connection
from leader to follower potentiates whereas the other depresses.
Finally in the vicinity of anti-phase forcing both synapses depress
and the neurons  decouple. Analysis of Eqs.\ref{eq:ws} furthermore
reveals several regions of bistability  between these different
motifs. Simulations agree well with the analysis, see
Fig.\ref{fig:traces}. Previous work on STDP in a network model of 
hippocampus found similar effects of the phase of oscillation on 
connectivity motifs through simulation \citep{bush10}.

The evolution equations for the synaptic weights of a network of
arbitrary size can be derived  using the separation of time scales
technique. However, for several cases of interest the  theory for
pairs of neurons can be used to gain insight into the resutling
connectivity of large  networks. This includes the case of several
clusters forced at different phases, as well as the  case of a uniform
distribution of phases. When phases are widely distributed the general
finding is  the emergence of hierarchical structure. Specifically,
there is clustering locally between neurons  with similar phases of
the forcing, but between neuron pairs with disparate phases,
unidirectional  connections form according to the leader-follower
phase-relationship. It is interesting to note that  precisely this
type of hierarchical clustering in cortical microcircuits has been
inferred from  data collected through multiple patch-clamp experiments
in slices \citep{vegue17}.  

\textit{Noisy drive:} In the case of two neurons forced by white-noise
inputs with a given  correlation and time-delay, the resulting phase
diagram is very rich, see Fig.\ref{fig:phaseplanes_noise}.  Although
positive correlations tend to lead to potentiation and negative
correlations to  depression, the combined effect of correlation and
delay is complex, and multistability is the  rule. This is borne out
in numerical simulations for pairs of neurons, see
Fig.\ref{fig:phaseplane_w_traces}.  It is unclear how this will affect
the emergent connectivity in large networks of neurons, and  requires
additional study. 

The plasticity process we describe here consists of a build up over
time of a large  number of small changes. As such, it is slow, and
would not be a relevant  mechanism for rapid memory formation, such as
episodic memory. Rather, such a  process shapes the connectivity in
recurrent circuits in accordance with  regularities in the statistics
of the inputs. For example, if we consider an  area in the visual
pathway, neurons with similar  feature selectivity and overlapping
receptive fields would exhibit positive correlations  in their output
in response  to a time-varying stimulus. Both for the case of
oscillatory as well as noisy  dynamics our analysis would predict a
potentiation of recurrent connections  between these neurons. This
would lead to an enhanced response to a similar  stimulus over
time. Neurons with similar feature selectivity but non-overlapping
receptive fields would likely exhibit similar yet time-delayed (or
out-of-phase) inputs  in response to the motion of an object across
the visual field, etc.  Therefore, we expect that the statistics of
some sensory stimuli can be mapped on to the parameters of the inputs
in our model: frequency, phase, input-correlation, delay. Repeated exposure 
to the same sensory stimuli would lead to a slow reshaping of the recurrent 
circuit connectivity and hence the neuronal response. Such a process may be 
relevant for the phenomenon of perceptual learning \citep{gilbert01}.

\section*{Appendix A: Derivation of self-consistent equations for synaptic weights}
\label{sec:appendix_A}
Here we derive a self-consistent set of equations for the synaptic weights $w_{12}$ and $w_{21}$ by assuming a 
separation of time-scales between the rate dynamics and the synaptic plasticity. We first note that the 
evolution equations for the synaptic weights, Eqs.\ref{eq:wij_ns} can be rewritten for the case of 
stationary rate dynamics by noting that $\langle r_{i}(t)r_{j}(t-T)\rangle_{t} = \langle r_{i}(t+T)r_{j}(t)\rangle_{t}$, which 
allows for a change of variables in the second integral, leading to Eq.\ref{eq:wij}. Strictly speaking this 
correspondence only holds when the dynamics has been averaged over the fast time. As we are only 
interested in the slow-time dynamics, we write Eq.\ref{eq:wij} as if the correspondence were exact, 
cognizant that it is a slight abuse of notation.

Now, given real-valued, time-varying 
inputs to the two neurons, the equations are 
\begin{equation}
\begin{aligned}
\tau\dot{r}_{1} &= -r_{1}+w_{12}r_{2}+I_{1}(t),\\
\tau\dot{r}_{2} &= -r_{2}+w_{21}r_{1}+I_{2}(t),\\
\dot{w}_{12} &= \int_{-\infty}^{\infty}dT A(T)r_{2}(t)r_{1}(t+T),\\
\dot{w}_{21} &= \int_{-\infty}^{\infty}dT A(T)r_{1}(t)r_{2}(t+T),\label{eqs:rate_weights}
\end{aligned}
\end{equation}
where $A(T)$ is the plasticity rule.  There is no general analytical 
solution to these equations given the quadratic nonlinearities.  
However, if we assume that each
synaptic weight change is small, then the synaptic weights will
evolve much more slowly than the rates and we can formally separate
the time scales in a multi-scale analysis.  To do this we introduce
the small parameter $\epsilon\ll 1$ such that 
$A(T) = \epsilon\tilde{A}(T)$.  We also introduce the slow time $t_{s} = \epsilon t$
and allow for the rates and weights to  evolve on both fast and slow
time scales, i.e. they are functions of $t$ and $t_{s}$, and these two
times are taken to be independent variables. 

Then we can write
\begin{equation}
\begin{aligned}
r_{1} &= r_{1}^{0}(t,t_{s})+\epsilon
r_{1}^{1}(t,t_{s})+\mathcal{O}(\epsilon^{2}),\\ r_{2} &=
r_{2}^{0}(t,t_{s})+\epsilon
r_{2}^{1}(t,t_{s})+\mathcal{O}(\epsilon^{2}),\\ w_{12} &=
w_{12}^{0}(t,t_{s})+\epsilon
w_{12}^{1}(t,t_{s})+\mathcal{O}(\epsilon^{2}),\\ w_{21} &=
w_{21}^{0}(t,t_{s})+\epsilon
w_{21}^{1}(t,t_{s})+\mathcal{O}(\epsilon^{2}).
\end{aligned}
\end{equation}
Pluggin these into Eqs.\ref{eqs:rate_weights} and collecting terms
order-by-order gives, at order $\mathcal{O}(1)$
\begin{equation}
\begin{aligned}
\tau\partial_{t}r_{1}^{0} &=
-r_{1}^{0}+w_{12}^{0}r_{2}^{0}+I_{1}(t),\\ \tau\partial_{t}r_{2}^{0}
&= -r_{2}^{0}+w_{21}^{0}r_{1}^{0}+I_{2}(t),\\ \partial_{t}w_{12}^{0}
&= 0,\\ \partial_{t}w_{21}^{0} &= 0.
\end{aligned}
\end{equation}
These last two equations show that the leading-order weights only
depend on the slow time, namely $w_{12}^{0}(t,t_{s}) =
w_{12}^{0}(t_{s})$ and $w_{21}^{0}(t,t_{s}) = w_{21}^{0}(t_{s})$.
Therefore, they can be treated as constants in  the rate equations,
which evolve on the fast time-scale. 

At order $\mathcal{O}(\epsilon )$ we have
\begin{equation}
\begin{aligned}
\tau\partial_{t_{s}}r_{1}^{0}+\tau\partial_{t}r_{1}^{1} &= -r_{1}^{1}+w_{12}^{1}r_{2}^{0}+w_{12}^{0}r_{2}^{1},\\
\tau\partial_{t_{s}}r_{2}^{0}+\tau\partial_{t}r_{2}^{1} &= -r_{2}^{1}+w_{21}^{1}r_{1}^{0}+w_{21}^{0}r_{1}^{1},\\
\partial_{t_{s}}w_{12}^{0}+\partial_{t}w_{12}^{1} &= \int\limits_{-\infty}^{\infty}dT\tilde{A}(T)r_{2}^{0}(t)r_{1}^{0}(t+T),\\
\partial_{t_{s}}w_{21}^{0}+\partial_{t}w_{21}^{1} &= \int\limits_{-\infty}^{\infty}dT\tilde{A}(T)r_{1}^{0}(t)r_{2}^{0}(t+T).
\end{aligned}
\end{equation}
The first two equations give a correction to the leading order
solution of the firing rates, which we will not use here.  The weight
equations at first  glance do not seem solvable since we are expected
to solve for both the leading order solution of the synaptic weights
as well as the next order  correction in the same set of equations.
However, we know that the leading order terms are independent of the
fast time $t$, which will allow us to solve for both. Specifically,
the evolution of the leading-order weights  will depend only on those
terms from the integral which are independent of the  fast time. This
leads to Eqs.\ref{eqs:asymp}. For simplicity in notation, in what
follows we will drop the  superscripts and tildes and write, for the
leading-order solution, simply
\begin{equation}
\begin{aligned}
\tau\partial_{t}r_{1} &=
-r_{1}+w_{12}r_{2}+I_{1}(t),\\ \tau\partial_{t}r_{2}
&=
-r_{2}+w_{21}r_{1}+I_{2}(t),\\ \partial_{t_{s}}w_{12}
&= \int
dt\int_{-\infty}^{\infty}dTA(T)r_{2}(t)r_{1}(t+T),\\ \partial_{t_{s}}w_{21}
&= \int
dt\int_{-\infty}^{\infty}dTA(T)r_{1}(t)r_{2}(t+T).\label{eqs:asymp}
\end{aligned}
\end{equation}
We will consider the specific cases of oscillatory and noisy drive below. 

\section*{Appendix B: Oscillatory drive}
\label{sec:appendix_B}

Here we study the case where the neurons are driven sinusoidally with
a frequency $\omega$ and with a phase  difference of $\phi$,
i.e. $I_{1} = I_{0}+Ie^{i\omega t+i\phi_{1}}$ and $I_{2} = I_{0}+Ie^{i\omega
  t+i\phi_{2}}$. The (complex) rates can be  written $(r_{1},r_{2}) =
(R_{10}(t_{s}),R_{20}(t_{s}))+(R_{11}(t_{s}),R_{21}(t_{s}))e^{i\omega
  t}$. We find that 
\begin{equation}
\begin{aligned}
R_{10} &= I_{0}\frac{(1+w_{12})}{1-w_{12}w_{21}},\\ R_{20} &=
I_{0}\frac{(1+w_{21})}{1-w_{12}w_{21}},\\ R_{11} &=
R\Big((1+i\tau\omega)e^{i\phi_{1}} +w_{12}^{0}e^{i\phi_{2}}\Big),\\ R_{21} &=
R\Big(w_{21}e^{i\phi_{1}}+(1+i\tau\omega )e^{i\phi_{2}}\Big),\\ R &=
\frac{I}{(1+i\tau\omega)^{2}-w_{12}^{0}w_{21}^{0}}.\label{eqs:pairwise_amp}
\end{aligned}
\end{equation}
Because we only consider balanced plasticity rules here, the constant,
baseline rates will not affect the  synaptic rates. Nonetheless, when
conducting numerical simulations it is important to take large enough
constant drive $I_{0}$ to ensure positive rates. 

In order to calculate the equations for the synaptic weights we must
use the real part of the  complex rates. As an illustration we
consider the equation for the weight $w_{12}$, which is
\begin{equation}
\partial_{t_{s}}w_{12} = \int
dt\int_{-\infty}^{\infty}dT\bar{A}(T)\mathrm{Re}(r_{2}(t))\mathrm{Re}(r_{1}(t+T)).\nonumber
\end{equation}
The quadratic term 
\begin{eqnarray}
&&\mathrm{Re}(r_{2}(t))\mathrm{Re}(r_{1}(t+T)) = \frac{1}{4}\Big(R_{2}e^{i\omega t}+\bar{R}_{2}e^{-i\omega t}\Big)
\nonumber\\&&\cdot\Big(R_{1}e^{i\omega t+i\omega T}+\bar{R}_{1}e^{-i\omega t-i\omega T}\Big),\nonumber\\
&=& \frac{1}{4}\Big(R_{1}\bar{R}_{2}e^{i\omega T}+\bar{R}_{1}R_{2}e^{-i\omega T}\Big)
\nonumber\\&&+\frac{1}{4}\Big(R_{1}R_{2}e^{2i\omega t+i\omega T}+\bar{R}_{1}\bar{R}_{2}e^{-2i\omega t-i\omega T}\Big)\nonumber.
\end{eqnarray}
Note that the first two terms are independent of the fast time $t$,
while the second two terms oscillate  on the fast timescale with a
frequency $2\omega$. Integrating over the fast timescale therefore
eliminates the  latter terms. Performing the second integral and then
doing the analogous calculation for the other weight leads to
Eqs.\ref{eq:ws} where $\phi = \phi_{2}-\phi_{1}$.

\noindent
\textit{Growth rate of synaptic weights}. While the final state of the synaptic weights depends on the 
phase difference, the rate at which plasticity occurs is strongly influenced by the 
frequency of forcing. This can be most easily seen for the case of in-phase forcing $\phi = 0$, for which 
we expect both weights to potentiate (or depress for an anti-hebbian rule). Assuming $w_{ij} = w_{ji}$ leads to 
a right-hand side (growth rate) of Eqs.\ref{eq:ws} which is simply proportional to 
$\tilde{A}_{+}(\omega )-\tilde{A}_{-}(\omega )$ which is zero for $\omega = 0$ and 
as $\omega\to\infty$, while it has a maximum for $\omega = 1/\sqrt{\tau_{+}\tau_{-}}$.

\subsection*{Theory for networks}
For the case of $n$ coupled neurons, the rate equation for the $i$th neuron is 
\begin{equation}
\tau r_{i} = -r_{i}+\frac{1}{N}\sum_{j=1}^{N}w_{ij}r_{j}+I_{i},\label{eqs:ri}
\end{equation}
where $I_{i} = I_{0}+Ie^{i\phi_{i}}$, while the evolution equation for the synaptic weight from neuron $j$ to neuron $i$ is still 
described by Eq.\ref{eq:wij}. We can once again apply the separation of timescales formally 
by defining $A(T) = \epsilon\tilde{A}(T)$ where $\epsilon\ll 1$ and defining the slow time 
$t_{s} = \epsilon t$. The rates can be written in vector form as 
$\mathbf{r}(t,t_{s}) = \mathbf{R}_{0}(t_{s})+\mathbf{R}_{1}(t_{s},\omega)e^{i\omega t}$ where 
\begin{eqnarray}
\mathbf{R}_{0} &=& I_{0}\Big(\mathbf{I}-\mathbf{W}\Big)^{-1}\mathbf{e},\nonumber\\
\mathbf{R}_{1} &=& I\Big((i\tau\omega +1)\mathbf{I}-\mathbf{W}\Big)^{-1}\mathbf{p},
\end{eqnarray}
where $\mathbf{I}$, $\mathbf{W}$ are the identity matrix and weight matrix respectively, 
$\mathbf{e}$ is a vector of ones, while the 
$j$th element of the vector $\mathbf{p}$ is $e^{i\phi_{j}}$.

\noindent
\textit{Applicability of pairwise theory to network simulations}
If we consider the rate equations for a pair of neurons $j$ and $k$ in the network Eqs.\ref{eqs:ri} we find, applying the 
separation of time-scales approach detailed in Appendix A, that 
the oscillatory components obey 
\begin{eqnarray}
\tau\dot{R}_{j1} &=& -R_{j1}+\frac{1}{N}w_{jk}^{0}R_{k1}+Ie^{i\phi_{j}}+\xi_{j},\nonumber\\
\tau\dot{R}_{k1} &=& -R_{k1}+\frac{1}{N}w_{kj}^{0}R_{j1}+Ie^{j\phi_{k}}+\xi_{k},
\end{eqnarray}
where $\xi_{a} = \frac{1}{N}\sum_{l=1,\ne j,k}^{N}w_{al}^{0}R_{l1}$. From this we find the complex amplitudes
\begin{eqnarray}
R_{j1} &=& R\Big((1+i\tau\omega)(Ie^{i\phi_{j}}+\xi_{j})+\frac{1}{N}w_{jk}^{0}(Ie^{i\phi_{k}}+\xi_{k})\Big),\nonumber\\
R_{k1} &=& R\Big(\frac{1}{N}w_{kj}^{0}(Ie^{i\phi_{j}}+\xi_{j})+(1+i\tau\omega )(Ie^{i\phi_{k}}+\xi_{k})\Big),
\end{eqnarray}
where $R = 1/((1+i\tau\omega)^{2}-w_{jk}^{0}w_{kj}^{0})$. Note that these equations are identical to 
those for the complex amplitudes for the pairwise case (with renormalized weights), Eqs.\ref{eqs:pairwise_amp}, with the exception of 
the meanfield terms $\xi_{j}$ and $\xi_{k}$. The slow dynamics of the synaptic weight $w_{jk}$ is then given by
\begin{equation}
\partial_{t_{s}}w_{jk} = \int
dt\int_{-\infty}^{\infty}dT\bar{A}(T)\mathrm{Re}(r_{k}(t))\mathrm{Re}(r_{j}(t+T)).\nonumber
\end{equation}
Note that in principle the rates $r_{k}$ and $r_{j}$ still depend on the meanfield terms and hence this equation 
is not self-consistent as in the pairwise case. The
influence of these meanfield terms depends strongly on the distribution of phases of the complex amplitudes. 
In one of the two limiting cases, if all of the phases are aligned then the moduli of the terms all sum. This is 
equivalent to the summation of vectors all with the same angle. In the other limiting case, if the phases are 
uniformaly distributed, then the resultant modulus will be close to zero because we are summing many vectors 
all with distinct phases (as long as the moduli and phases are only weakly correlated or uncorrelated). Hence in 
this limit the inlfuence of the meanfield vanishes and only the pairwise interactions matter. This latter 
case is the relevant one for Fig.\ref{fig:networks}C and explains why the pairwise theory correctly predicts the 
network structure after learning. 
 
\subsection*{Network simulations}
For the simulations shown in Fig.\ref{fig:two_clusters}, the following nonlinear rate equations were used
\begin{equation}
\tau\dot{\mathbf{r}} = -\mathbf{r}+\alpha\Theta\Big(\frac{1}{K}\mathbf{W}\mathbf{r}-I_{0}+I\Big)\nonumber\\,
\end{equation} 
where $\Theta$ is the Heaviside function. A spike from a neuron $i$ in a timestep $\Delta t$ occurs 
with probability $r_{i}\Delta t$. Given the spike trains from neurons $i$ and $j$, a weight 
$w_{ij}$ undergoes updates from all spike pairs according to the STDP rule, see \citep{pfister06} for 
the numerical scheme. For the simulations in Fig.\ref{fig:networks} linear rate equations are used.

\section*{Appendix C: Noisy drive}
\label{sec:appendix_C}
Here we consider an external drive of the form
\begin{equation}
\begin{aligned}
I_{1}(t) &= \sqrt{1-|c|}\xi_{1}(t)+\mathrm{sgn}(c)\sqrt{|c|}\xi_{c}(t),\\
I_{2}(t) &= \sqrt{1-|c|}\xi_{2}(t)+\sqrt{|c|}\xi_{c}(t-D),
\end{aligned}
\end{equation}
where $\xi_{i}(t)$ is a Gaussian white noise process,
i.e. $\langle\xi_{i}(t)\rangle = 0$ and
$\langle\xi_{i}(t)\xi_{j}(t^{'})\rangle = \sigma^{2}\delta_{ij}\delta(t-t^{'})$. 
Therefore, the noisy drive to the two neurons has correlation $c$. The correlated 
input is delayed to neuron 2 with respect to neuron 1 by a time $D$.

To solve the system of rate equations we rewrite it in
vector form as 
\begin{equation}
\tau \mathbf{\dot{r}} = \mathbf{W}\mathbf{r}+\mathbf{I},
\end{equation}
where $\mathbf{r} = (r_{1},r_{2})$, $\mathbf{I} = (I_{1},I_{2})$ and 
\begin{equation}
\mathbf{W} = 
\left(
\begin{array}{cc}
-1 & w_{12}\\
w_{21} & -1\\
\end{array}\right).
\end{equation}
We diagonalize the connectivity matrix $\mathbf{W} = \mathbf{Q}\mathbf{\Lambda}\mathbf{Q}^{-1}$ and obtain the system of independent equations
\begin{equation}
\tau\mathbf{\dot{u}} = \mathbf{\Lambda}\mathbf{u}+\mathbf{Q}^{-1}\mathbf{I},
\end{equation}
where $\mathbf{u} = \mathbf{Q}^{-1}\mathbf{r}$.  The matrices resulting from the diagonalization are 
\begin{eqnarray}
\mathbf{Q} &=& 
\left(
\begin{array}{cc}
-\sqrt{w_{12}/w_{21}} & \sqrt{w_{12}/w_{21}} \\
1 & 1\\
\end{array}
\right),
\mathbf{\Lambda} = 
\left(
\begin{array}{cc}
-1-\sqrt{w_{12}w_{21}} & 0 \\
0 & -1+\sqrt{w_{12}w_{21}}\\
\end{array}
\right),\nonumber\\
\mathbf{Q}^{-1} &=& 
\left(
\begin{array}{cc}
-1 & \sqrt{w_{12}/w_{21}} \\
1 & \sqrt{w_{12}/w_{21}}\\
\end{array}
\right).
\end{eqnarray}

The equations for the transformed variables $\mathbf{u}$ are 
\begin{eqnarray}
\tau\dot{u}_{1} &=& -(1+\sqrt{w_{12}w_{21}})u_{1}+\frac{1}{2}\Big(-\sqrt{\frac{w_{21}}{w_{12}}}I_{1}(t)+I_{2}(t)\Big),\\
\tau\dot{u}_{2} &=& -(1-\sqrt{w_{12}w_{21}})u_{1}+\frac{1}{2}\Big(\sqrt{\frac{w_{21}}{w_{12}}}I_{1}(t)+I_{2}(t)\Big).
\end{eqnarray}
These equations can be solved formally as
\begin{equation}
\begin{aligned}
u_{1}(t) &= u_{1}(0)e^{-t/\tau_{f}}-\sqrt{1-|c|}\frac{\sigma}{2\sqrt{\tau}}\sqrt{\frac{w_{21}}{w_{12}}}\int\limits_{0}^{t}e^{-(t-u)/\tau_{f}}dW_{u}+\sqrt{1-|c|}\frac{\sigma}{2\sqrt{\tau}}\int\limits_{0}^{t}e^{-(t-v)/\tau_{f}}dW_{v}
\\&-\frac{\sigma}{2\sqrt{\tau}}\sqrt{\frac{w_{21}}{w_{12}}}\mathrm{sgn}(c)\sqrt{|c|}\int\limits_{0}^{t}dW_{r}e^{-(t-r)/\tau_{f}}
+\frac{\sigma}{2\sqrt{\tau}}\sqrt{|c|}\int_{0}^{t-D}dW_{r}e^{-(t-r)/\tau_{f}},\\
u_{2}(t) &= u_{2}(0)e^{-t/\tau_{s}}+\sqrt{1-|c|}\frac{\sigma}{2\sqrt{\tau}}\sqrt{\frac{w_{21}}{w_{12}}}\int\limits_{0}^{t}e^{-(t-u)/\tau_{s}}dW_{u}+\sqrt{1-|c|}\frac{\sigma}{2\sqrt{\tau}}\int\limits_{0}^{t}e^{-(t-v)/\tau_{s}}dW_{v}
\\&+\frac{\sigma}{2\sqrt{\tau}}\sqrt{\frac{w_{21}}{w_{12}}}\mathrm{sgn}(c)\sqrt{|c|}\int\limits_{0}^{t}dW_{r}e^{-(t-r)/\tau_{s}}
+\frac{\sigma}{2\sqrt{\tau}}\sqrt{|c|}\int_{0}^{t-D}dW_{r}e^{-(t-r)/\tau_{s}}
\end{aligned}
\end{equation}
where $dW_{u}$, $dW_{v}$ and $dW_{r}$ are the stochastic differentials
corresponding to the Gaussian processes $\xi_{1}$, $\xi_{2}$ and
$\xi_{c}$ respectively.  Also, we have defined the fast and slow time
constants 
\begin{equation}
\begin{aligned}
\tau_{f} &= \frac{\tau}{1+\sqrt{w_{12}w_{21}}},\\
\tau_{s} &= \frac{\tau}{1-\sqrt{w_{12}w_{21}}},
\end{aligned}
\end{equation}
from which it is clear that there is an instability for
$w_{12}w_{21}>1$.  The original firing rates are linear combinations
of these variables.  Specifically,
\begin{equation}
\begin{aligned}
r_{1} &= \sqrt{\frac{w_{12}}{w_{21}}}(-u_{1}+u_{2}),\\
r_{2} &= u_{1}+u_{2}.
\end{aligned}
\end{equation}
Finally, we have, and ignoring the dependence on the initial condition,
\begin{equation}
\begin{aligned}
r_{1}(t) &= \sqrt{1-|c|}\frac{\sigma}{2\sqrt{\tau}}\Bigg[\int\limits_{0}^{t}\Big(e^{-(t-u)/\tau_{f}}+e^{-(t-u)/\tau_{s}}\Big)dW_{u}\\
&+\sqrt{\frac{w_{12}}{w_{21}}}\int\limits_{0}^{t}\Big(e^{-(t-v)/\tau_{s}}-e^{-(t-u)/\tau_{f}}\Big)dW_{v}\Bigg]\\
&+\sqrt{|c|}\frac{\sigma}{2\sqrt{\tau}}\Bigg[\mathrm{sgn}(c)\int\limits_{0}^{t}dW_{r}e^{-(t-r)/\tau_{f}}
+\mathrm{sgn}(c)\int\limits_{0}^{t}dW_{r}e^{-(t-r)/\tau_{s}}\\
&-\sqrt{\frac{w_{12}}{w_{21}}}\int\limits_{0}^{t-D}dW_{r}e^{-(t-r)/\tau_{f}}+\sqrt{\frac{w_{12}}{w_{21}}}\int\limits_{0}^{t-D}dW_{r}e^{-(t-r)/\tau_{s}}\Bigg],\\
r_{2}(t) &= \sqrt{1-|c|}\frac{\sigma_{1}}{2\sqrt{\tau}}\sqrt{\frac{w_{21}}{w_{12}}}\int\limits_{0}^{t}\Big(e^{-(t-u)/\tau_{s}}-e^{-(t-u)/\tau_{f}}\Big)dW_{u}\\&+\sqrt{1-|c|}\frac{\sigma_{2}}
{2\sqrt{\tau}}\int\limits_{0}^{t}\Big(e^{-(t-v)/\tau_{s}}+e^{-(t-u)/\tau_{f}}\Big)dW_{v}\\
&+\sqrt{|c|}\frac{\sigma_{c}}{2\sqrt{\tau}}\Bigg[-\mathrm{sgn}(c)\sqrt{\frac{w_{21}}{w_{12}}}\int\limits_{0}^{t}dW_{r}e^{-(t-r)/\tau_{f}}
+\mathrm{sgn}(c)\sqrt{\frac{w_{21}}{w_{12}}}\int\limits_{0}^{t}dW_{r}e^{-(t-r)/\tau_{s}}\\
&+\int\limits_{0}^{t-D}dW_{r}e^{-(t-r)/\tau_{f}}+\int\limits_{0}^{t-D}dW_{r}e^{-(t-r)/\tau_{s}}\Bigg].
\end{aligned}
\end{equation}

The slow dynamics of the synaptic weights, which is calculated
self-consistently through the rates, is therefore also  stochastic. In
this case the integral over the fast time in Eqs.\ref{eqs:asymp}
yields the expected value of the  product of rates. Namely,
\begin{equation}
\partial_{t_{s}}w_{12} = \int\limits_{-\infty}^{\infty}dTA(T)E(r_{2}(t)r_{1}(t+T)),\nonumber
\end{equation}
and similarly for $w_{21}$. Evaluating this expectation requires
products of stochastic integrals. For independent  processes this
expectation is always zero, while for integrals of the same process
the product can be expressed  as a standard integral through the
so-called Ito isometry. For example,
\begin{eqnarray}
E\Bigg(\int\limits_{0}^{t}e^{-\frac{(t-u)}{\tau}}dW_{u}\int\limits_{0}^{t+T}e^{-\frac{(t+T-u)}{\tau}}dW_{u}\Bigg) 
&=& e^{-\frac{(2t+T)}{\tau}}E\Bigg(\int\limits_{0}^{t}e^{\frac{u}{\tau}}dW_{u}\int\limits_{0}^{t+T}e^{\frac{u}{\tau}}dW_{u}\Bigg)\nonumber\\
&=& e^{-\frac{(2t+T)}{\tau}}\int\limits_{0}^{\mathrm{min}(t,t+T)}e^{\frac{2u}{\tau}}du,\nonumber\\
&=& 
\begin{cases}
e^{-\frac{(2t+T)}{\tau}}\frac{\tau}{2}\Big(e^{\frac{2t}{\tau}}-1\Big) & T>0\\
e^{-\frac{(2t+T)}{\tau}}\frac{\tau}{2}\Big(e^{\frac{2(t+T)}{\tau}}-1\Big) & T\le 0
\end{cases}\nonumber\\
&=& \frac{\tau}{2}\Big(e^{-\frac{|T|}{\tau}}-e^{-\frac{(2t+T)}{\tau}}\Big),
\end{eqnarray}
which is independent of $t$ at long times.
Performing these integrals yields the evolution equations, Eqs.\ref{eq:ws_noise}.

\subsection*{Evolution equations for small weights} 
If $w_{12}\ll 1$ and $w_{21}\ll 1$ then
\begin{eqnarray}
\dot{w}_{21} &=& \frac{c\sigma_{c}^{2}}{2}A_{+}\tau_{+}\tau\Big(\frac{1}{\tau+\tau_{+}}-\frac{1}{\tau+\tau_{-}}\Big)\nonumber\\
&&+\frac{A_{+}\tau_{+}\tau}{4}\Bigg\{w_{21}\Big((1-|c|)\sigma_{1}^{2}+|c|\sigma_{c}^{2}\Big)\Big(\frac{2\tau_{+}}{(\tau+\tau_{+})^{2}}+\frac{1}{\tau+\tau_{+}}-\frac{1}{\tau+\tau_{-}}\Big)\nonumber\\
&&+w_{12}\Big((1-|c|)\sigma_{2}^{2}+|c|\sigma_{c}^{2}\Big)\Big(\frac{-2\tau_{-}}{(\tau+\tau_{-})^{2}}+\frac{1}{\tau+\tau_{+}}-\frac{1}{\tau+\tau_{-}}\Big)\Bigg\},\nonumber\\
\dot{w}_{12} &=& \frac{c\sigma_{c}^{2}}{2}A_{+}\tau_{+}\tau\Big(\frac{1}{\tau+\tau_{+}}-\frac{1}{\tau+\tau_{-}}\Big)\nonumber\\
&&+\frac{A_{+}\tau_{+}\tau}{4}\Bigg\{w_{12}\Big((1-|c|)\sigma_{2}^{2}+|c|\sigma_{c}^{2}\Big)\Big(\frac{2\tau_{+}}{(\tau+\tau_{+})^{2}}+\frac{1}{\tau+\tau_{+}}-\frac{1}{\tau+\tau_{-}}\Big)\nonumber\\
&&+w_{21}\Big((1-|c|)\sigma_{1}^{2}+|c|\sigma_{c}^{2}\Big)\Big(\frac{-2\tau_{-}}{(\tau+\tau_{-})^{2}}+\frac{1}{\tau+\tau_{+}}-\frac{1}{\tau+\tau_{-}}\Big)\Bigg\}.\nonumber
\end{eqnarray}

If $w_{12}\ll 1$ and $w_{21}$ can be order one, then
\begin{eqnarray}
\dot{w}_{21} &=& \frac{c\sigma_{c}^{2}}{2}A_{+}\tau_{+}\tau\Big(\frac{1}{\tau+\tau_{+}}-\frac{1}{\tau+\tau_{-}}\Big)\nonumber\\
&&+\frac{A_{+}\tau_{+}\tau}{4}w_{21}\Big((1-|c|)\sigma_{1}^{2}+|c|\sigma_{c}^{2}\Big)\Big(\frac{2\tau_{+}}{(\tau+\tau_{+})^{2}}+\frac{1}{\tau+\tau_{+}}-\frac{1}{\tau+\tau_{-}}\Big),\nonumber\\
\dot{w}_{12} &=& \frac{c\sigma_{c}^{2}}{2}A_{+}\tau_{+}\tau\Big(\frac{1}{\tau+\tau_{+}}-\frac{1}{\tau+\tau_{-}}\Big)\nonumber\\
&&+\frac{A_{+}\tau_{+}\tau}{4}w_{21}\Big((1-|c|)\sigma_{1}^{2}+|c|\sigma_{c}^{2}\Big)\Big(\frac{-2\tau_{-}}{(\tau+\tau_{-})^{2}}+\frac{1}{\tau+\tau_{+}}-\frac{1}{\tau+\tau_{-}}\Big).\nonumber
\end{eqnarray}

If $w_{21}\ll 1$ and $w_{12}$ can be order one, then
\begin{eqnarray}
\dot{w}_{21} &=& \frac{c\sigma_{c}^{2}}{2}A_{+}\tau_{+}\tau\Big(\frac{1}{\tau+\tau_{+}}-\frac{1}{\tau+\tau_{-}}\Big)\nonumber\\
&&+\frac{A_{+}\tau_{+}\tau}{4}w_{12}\Big((1-|c|)\sigma_{2}^{2}+|c|\sigma_{c}^{2}\Big)\Big(\frac{-2\tau_{-}}{(\tau+\tau_{-})^{2}}+\frac{1}{\tau+\tau_{+}}-\frac{1}{\tau+\tau_{-}}\Big),\nonumber\\
\dot{w}_{12} &=& \frac{c\sigma_{c}^{2}}{2}A_{+}\tau_{+}\tau\Big(\frac{1}{\tau+\tau_{+}}-\frac{1}{\tau+\tau_{-}}\Big)\nonumber\\
&&+\frac{A_{+}\tau_{+}\tau}{4}w_{12}\Big((1-|c|)\sigma_{2}^{2}+|c|\sigma_{c}^{2}\Big)\Big(\frac{2\tau_{+}}{(\tau+\tau_{+})^{2}}+\frac{1}{\tau+\tau_{+}}-\frac{1}{\tau+\tau_{-}}\Big).\nonumber
\end{eqnarray}

\begin{acknowledgements}
AR acknowledges ``Retos'' project RTI2018-097570-B-100 from the Ministry of Science and Innovation of the Spanish 
Government, Flag-Era project from the EU for the Human Brain Project HIPPOPLAST (Era-ICT code 
PCI2018-093095), ``Red de Investigaci\'on'' RED2018-102323-T from the Ministry of Science and Innovation of the 
Spanish Government. This work is supported by the Spanish State Research Agency, through the Severo Ochoa and 
Maria de Maeztu program for Centers and Units of Excellence in R\&D (CEX2020-001084-M). We thank CERCA 
Program/Generalitat de Catalunya for institutional support. 
We acknowledge very helpful discussions with Marina Vegu\'e, Toni Guillamon and Ernest 
Montrbri\'o.
\end{acknowledgements}

\bibliographystyle{spbasic}

\end{document}